# Propagation of Ultra High Energy Protons over Cosmological Distances and Implications for Topological Defect Models

R. J. Protheroe and P. A. Johnson
Department of Physics and Mathematical Physics
The University of Adelaide, Adelaide, Australia 5005.

**Abstract**

We describe the results of a hybrid matrix–Monte Carlo calculation of cascading of UHE cosmic rays and $\gamma$-rays through the cosmic background radiation fields over cosmological distances. We calculate the $\gamma$-ray and neutrino emission that results from the cascade, as well as the effect of cascading on the primary spectrum. We discuss the results for various cosmic ray injection spectra and primary species. Certain models for the production of the highest energy cosmic rays are ruled out.

## 1 Introduction

It was first noted by Greisen [1], and Zatsepin and Kuz'min [2] (GZK) that the nucleonic component of UHE cosmic rays above $10^{20}$ eV will be severely attenuated in the cosmic microwave background primarily due to pion photoproduction interactions with the low energy photons. Other processes such as nuclear disintegration [3, 4] and pair-production [5] have since been considered. A general consensus has emerged that no particles should be detected above the GZK cut-off if they are produced at moderate extragalactic distances. Accompanying the absorption will be a pile-up in the spectrum below the cut-off [6, 7], and this feature may well be present in the observed spectrum. There has been much debate about whether or not the observed cosmic ray spectrum cuts off sharply at $\sim 10^{20}$ eV [8, 9], but the recent detection of the two highest energy events at $3 \times 10^{20}$ eV [10] and $2 \times 10^{20}$ eV [11, 12], bringing to a total of 8 the number of cosmic rays detected above $10^{20}$ eV [13], dramatically confirms that the cosmic-ray spectrum does not end with the GZK cut-off. The more energetic of the two particles in particular, very likely could not have travelled more than 50 Mpc from its source if it was nucleonic or a $\gamma$-ray [14].

Monte Carlo [15, 16] and analytical [17] studies of the propagation have been performed in which the sensitivity of the spectrum to various source scenarios is investigated.

Here we consider in detail the propagation of UHE cosmic rays through the extragalactic background radiation field, including the cascade initiated by interactions of cosmic rays in the radiation. In this cascade we include $p\gamma$ pair-production and pion-production, $\gamma\gamma$ pair-production, inverse-Compton scattering and other processes, and use the matrix doubling method of Protheroe and Stanev [18] (see also [19]) to propagate the cascade up to distances equivalent to $z = 9$, and for primary energies up to $10^{23}$ eV. We calculate the accompanying $\gamma$-ray and neutrino fluxes in addition to the nucleon flux and examine several different source scenarios, investigating the properties of the cascade initiated by different types and energies of particles over different distances. Finally, we discuss the possibility that topological defects could be responsible for the "super-GZK events", i.e. cosmic rays observed above the expected GZK cut-off.

# 2 Interactions with Extragalactic Photons

## 2.1 The Extragalactic Photon Spectrum

The energy density of the extragalactic background radiation is dominated by that from the cosmic microwave background at a temperature of 2.735 K. Other components of the extragalactic background radiation are discussed in the review of Ressel and Turner [20]. The extragalactic radiation fields important for interactions of UHE cosmic rays include the cosmic microwave background, the radio background and the infrared–optical background. The radio background was measured over twenty years ago [21, 22], but the fraction of this radio background which is truly extragalactic, and not contamination from our own Galaxy, is still debatable. We use the original estimate [21], which is intermediate between other estimates [23, 14]. For the infrared and optical background, we use the model of Stecker *et al.* [24]. The composite spectrum used is given in Fig. 1.

## 2.2 Interactions and Energy-Loss Processes

There is a variety of energy-loss processes which are important for UHE cosmic rays propagating over large distances through a radiation field: protons interact with photons resulting in pion production and pair production; electrons interact via inverse-Compton scattering and triplet pair production, and emit synchrotron radiation in the intergalactic magnetic field; $\gamma$-rays interact by pair production. Energy losses due to cosmological redshifting of high energy particles and $\gamma$-rays are also important, and the cosmological redshifting of the background radiation fields means that energy thresholds and interaction lengths for the above processes also change with epoch.

We model photon-proton interactions as described in [25]. The mean interaction length, $\lambda$, of a proton of energy $E$ is given by,

$$[\lambda(E)]^{-1} = \frac{1}{8\beta E^2} \int_{\varepsilon_{\min}}^{\infty} \frac{n(\varepsilon)}{\varepsilon^2} \int_{s_{\min}}^{s_{\max}(\varepsilon,E)} \sigma(s)(s - m_p^2 c^4) ds d\varepsilon, \tag{1}$$

where $n(\varepsilon)$ is the differential photon number density of photons of energy $\varepsilon$, and $\sigma(s)$ is the appropriate total cross section for the process in question for a centre of momentum (CM) frame energy squared, $s$, given by

$$s = m_p^2 c^4 + 2\varepsilon E(1 - \beta \cos\theta) \tag{2}$$

where $\theta$ is the angle between the directions of the proton and photon, and $\beta c$ is the proton's velocity.

For pion photoproduction

$$s_{\rm min} = (m_p c^2 + m_\pi c^2)^2 \approx 1.16 \text{ GeV}^2, \tag{3}$$

and

$$\varepsilon_{\rm min} = \frac{m_\pi c^2(m_\pi c^2 + 2m_p c^2)}{2E(1+\beta)} \approx \frac{m_\pi c^2(m_\pi c^2 + 2m_p c^2)}{4E}. \tag{4}$$

For photon-proton pair-production the threshold is somewhat lower,

$$s_{\rm min} = (m_p c^2 + 2m_e c^2)^2 \approx 0.882 \text{ GeV}^2, \tag{5}$$

and

$$\varepsilon_{\rm min} \approx m_e c^2(m_e c^2 + m_p c^2)/E. \tag{6}$$

For both processes,

$$s_{\rm max}(\varepsilon, E) = m_p^2 c^4 + 2\varepsilon E(1+\beta) \approx m_p^2 c^4 + 4\varepsilon E, \tag{7}$$

and $s_{\rm max}(\varepsilon, E)$ corresponds to a head-on collision of a proton of energy $E$ and a photon of energy $\varepsilon$.

Examination of the integrand in Equation 1 shows that the energy of the soft photon interacting with a proton of energy $E$ is distributed as

$$p(\varepsilon) = \frac{\lambda(E) n(\varepsilon)}{8\beta E^2 \varepsilon^2} \Phi(s_{\rm max}(\varepsilon, E)) \tag{8}$$

in the range $\varepsilon_{\rm min} \leq \varepsilon \leq \infty$ where

$$\Phi(s_{\rm max}) = \int_{s_{\rm min}}^{s_{\rm max}} \sigma(s)(s - m_p^2 c^4) ds. \tag{9}$$

Similarly, examination of the integrand in Equation 1 shows that the square of the total CM frame energy is distributed as

$$p(s) = \frac{\sigma(s)(s - m_p^2 c^4)}{\Phi(s_{\rm max})}, \tag{10}$$

in the range $s_{\rm min} \leq s \leq s_{\rm max}$. The Monte Carlo rejection technique is used to sample $\varepsilon$ and $s$ respectively from the two distributions, and Equation 2 is used to find $\theta$. We then

Lorentz transform the interacting particles to the frame in which we treat the interaction (CM frame for pion photoproduction and proton rest frame for pair production), and sample momenta of particles produced in the interaction from the appropriate differential cross section by the rejection method. The energies of produced particles are then Lorentz transformed to the laboratory frame, and the final energy of the proton is obtained by requiring energy conservation. In this procedure, it is not always possible to achieve exact conservation of both momentum and energy while sampling particles from differential cross sections, and the momentum of the last particle sampled is therefore adjusted to minimize the error.

Following [25], for pion photoproduction we have used fits to the data given by Genzel, Joos and Pfeil [26] for the differential cross section for the two different channels for single pion production, and the inclusive data of Moffeit *et al.* [27] to sample the energies and momenta of the produced particles in the CM frame and Lorentz transform to the laboratory frame. For the total cross section at high energy we use ref. [28].

For photon-proton pair-production the threshold is somewhat lower than for pion photoproduction, and the total cross section for pair-production in the field of a proton is $\sim 100$ times larger than the total cross section for pion photoproduction. For the total cross section we use the Racah formula as parameterized by Maximon [29] (see Formula 3D-0000 in [30]). After sampling the energy and direction of the photon as described above and transforming to the proton rest frame, the positron energy is sampled from the single-differential cross section, $d\sigma/dE_+$ for which we use the Bethe-Heitler formula for an unscreened point nucleus (Formula 3D-1000 in [30]). Finally the positron's direction is sampled from the double-differential cross section, $d\sigma/dE_+d\Omega_+$ for which we use the Sauter-Gluckstern-Hull formula for an unscreened point nucleus (Formula 3D-2000 in [30]), and its laboratory frame energy is obtained by a Lorentz transformation. The cross section formulae neglect the recoil of the proton and so, as mentioned earlier, there is the problem of conserving both energy and momentum. Three choices present themselves: (a) conserve momentum in proton rest frame assuming no recoil $\vec{p}'_- = \vec{p}'_\gamma - \vec{p}'_+$; (b) ignore the electron, but set the distribution of electron lab frame energies to equal the distribution of positron energies, and use twice the positron's average lab frame energy for the average energy loss of the proton; (c) conserve energy in the proton rest frame assuming no recoil, $E'_- = E'_\gamma - E'_+$. We reject the last possibility because the electron's direction would have to be chosen arbitrarily before Lorentz transformation. In Figure 2 we compare the inelasticity,

$$\kappa(E) = 1 - \frac{\langle \Delta E \rangle}{E}, \tag{11}$$

calculated using the first two schemes. It is straightforward to show that the inelasticity at threshold must equal $2m_e/m_p \approx 10^{-3}$, and since scheme (b) is consistent with this we shall adopt it in our present work. The resulting inelasticity and energy-loss rate, obtained as described below, are in excellent agreement with those calculated analytically [5, 31, 32].

The mean interaction lengths for both processes, $\lambda(E)$, are obtained from Equation 1

for interactions in the microwave background and are plotted as dashed lines in Fig. 3. We use the inelasticity, $\kappa(E)$, to obtain the energy-loss distances for the two processes,

$$E/\frac{dE}{dx} = \frac{\lambda(E)}{\kappa(E)}. \tag{12}$$

Energy-loss distances are shown by the solid lines in Fig. 3, and in Fig. 4 we compare the total energy-loss distance from the present work with that of Berezinsky and Grigor'eva [7], Yoshida and Teshima [16] and Rachen and Biermann [32]. We are in excellent agreement with Rachen and Biermann in the region where pair production is dominant, but have a longer energy-loss distance for pion production, probably because of the different treatment of pion inelasticity. We agree better with Berezinsky and Grigor'eva over the decade in energy just above the threshold for pion production. Because cascading by pion photoproduction is rapid, and the proton energy is rapidly reduced to the GZK cut-off, minor differences in the pion photoproduction region will not have a great effect on predicted spectra except for nearby sources having spectra extending beyond 300 EeV. Overall, there is reasonable agreement between the four results below 300 EeV, except that the energy-loss distance calculated by Yoshida and Teshima is significantly above the other calculations at 50 EeV.

Inverse Compton interactions of high energy electrons and triplet pair production are modelled by the Monte Carlo technique as described in refs. [33, 34, 35, 36], and the mean interaction lengths and energy-loss distances for these processes are given in Fig. 5. Synchrotron losses are also included in our calculation and the energy-loss distance has been added to Fig. 5 for various magnetic fields.

Photon-photon interactions considered include pair production and double pair production [37], and the mean interaction lengths for these processes, showing the contribution of the various radiation fields, are illustrated in Fig. 6. Also shown is the mean interaction length for muon pair-production which is negligible in comparison with interactions with pair production on the radio background and double pair production on the microwave background.

# 3 The particle cascade

Where possible, to take account of the exact energy dependences of cross-sections, we use the Monte Carlo method. However, direct application of Monte Carlo techniques to cascades dominated by the physical processes described above over cosmological distances takes excessive computing time. The approach we use here is based on the matrix multiplication method described by Protheroe [33] and subsequently developed by Protheroe and Stanev [18]. We use a Monte Carlo program to calculate the yields of secondary particles due to interactions with radiation. Spectra of produced pions are decayed using routines in SIBYLL [38] to give yields of $\gamma, e, \nu_e, \bar{\nu}_e, \nu_\mu, \bar{\nu}_\mu$. Functions given in [34] are used to calculate synchrotron yields. The yields are then used to build up transfer matrices which describe the change in the spectra of particles produced after propagating through the radiation fields for a distance $\delta x$. Manipulation of the transfer matrices as described below enables one to calculate the spectra of particles resulting from propagation over arbitrarily large distances.

## 3.1 Matrix method

We use 180 fixed logarithmic energy bins of width $\Delta \log E = 0.1$ covering the energy range from $10^{-3}$ GeV to $10^{15}$ GeV. Thus, the energy range of the $j$th energy bin runs from $10^{(j-31)/10}$ GeV to $10^{(j-30)/10}$ GeV. The energy spectra of particles of type $\alpha$ ($\alpha = \gamma, e, p, n, \nu_e, \bar{\nu}_e, \nu_\mu, \bar{\nu}_\mu$) at distance $x$ in the cascade are represented by vectors $F_j^\alpha(x)$ which give the total number of particles of type $\alpha$ in the $j$th energy bin at distance $x$.

We define transfer matrices, $\mathrm{T}_{ij}^{\alpha\beta}(\delta x)$, which give the number of particles of type $\beta$ in the bin $j$ which result at a distance $\delta x$ after a particle of type $\alpha$ and energy in bin $i$ initiates a cascade. Then, given the spectra of particles at distance $x$ we can obtain the spectra at distance $(x + \delta x)$

$$F_j^\beta(x+\delta x) = \sum_\alpha \sum_{i=j}^{180} \mathrm{T}_{ij}^{\alpha\beta}(\delta x) F_i^\alpha(x) \tag{13}$$

where $F_i^\alpha(x)$ are the input spectra (number in the $i$th energy bin) of species $\alpha$.

We could also write this as

$$[\mathrm{F}(x+\delta x)] = [\mathrm{T}(\delta x)][\mathrm{F}(x)] \tag{14}$$

where

$$[\mathrm{F}] = \begin{bmatrix} F^\gamma \\ F^e \\ F^p \\ \vdots \end{bmatrix}, \quad [\mathrm{T}] = \begin{bmatrix} \mathrm{T}^{\gamma\gamma} & \mathrm{T}^{e\gamma} & \mathrm{T}^{p\gamma} & \cdots \\ \mathrm{T}^{\gamma e} & \mathrm{T}^{ee} & \mathrm{T}^{pe} & \cdots \\ \mathrm{T}^{\gamma p} & \mathrm{T}^{ep} & \mathrm{T}^{pp} & \cdots \\ \vdots & \vdots & \vdots & \ddots \end{bmatrix}. \tag{15}$$

## 3.2 Transfer matrix calculation

The transfer matrices depend on particle yields, $\mathrm{Y}_{\mathrm{ij}}^{\alpha\beta}$, which we define as the probability of producing a particle of type $\beta$ in the energy bin $j$ when a primary particle of type $\alpha$ with energy in bin $i$ undergoes an interaction. To calculate $\mathrm{Y}_{\mathrm{ij}}^{\alpha\beta}$ we use a Monte Carlo simulation. For inverse Compton scattering and photon-photon pair production we have used the computer code described by Protheroe [33, 34], updated to model interactions with a thermal photon distribution of arbitrary temperature, and arbitrary power-law spectra. For synchrotron radiation, we use the functions given in [34] to obtain the number of synchrotron photons produced in a given energy bin. To calculate the $\mathrm{T}_{\mathrm{ij}}^{\alpha\beta}$ we need both the mean interaction length of particles of type $\alpha$ and the distribution in energy per interaction of all produced particle types.

In the case of synchrotron radiation where the electron energy changes continuously at a rate

$$\frac{dE}{dx} = -bE^2 \tag{16}$$

we use, instead of the mean interaction length, the distance taken to lose energy from the centre of bin $i$ to the centre of bin $(i-1)$,

$$\lambda_{\rm eff}(E_i) = \frac{1}{b}\left(\frac{1}{E_{i-1}} - \frac{1}{E_i}\right), \tag{17}$$

and set $\mathrm{Y}^{\rm ee}_{\rm ij} = \delta_{\rm (i-1),j}$. We then use the functions given in [34] to obtain the yield, $\mathrm{Y}^{\rm e\gamma}_{\rm ij}$, i.e. the number of synchrotron photons produced in a given energy bin $j$ as the electron migrates from the centre of bin $i$ to the centre of bin $(i-1)$. The problem of continuous energy losses also applies to inverse Compton scattering in the Thomson regime and to proton-photon pair production, and we take the same approach in dealing with the problem. Formulae for the transfer matrices are given in Appendix A.

## 3.3 Matrix doubling

From Figs. 3 – 6 we see that at all epochs the smallest effective interaction length is that for synchrotron losses by electrons at high energies. We require $\delta x$ be much smaller than this distance which is of the order of parsecs for the highest magnetic field considered. The cascade is to be followed for a distance corresponding to a redshift of $z \sim 9$, and to complete the calculation of the cascade using repeated application of the transfer matrices would require $\sim 10^{12}$ steps. This is clearly impractical, and we must use the more sophisticated approach described below.

The matrix method and matrix doubling technique have been used for many years in radiative transfer problems [40, 41]. The method used here to calculate the spectrum of particles emerging after an arbitrary distance is that described by Protheroe and Stanev [18], and is summarized below. Once the transfer matrices have been calculated for a distance $\delta x$, the transfer matrix for a distance $2\delta x$ is simply given by applying the transfer matrices twice, i.e.

$$[\mathrm{T}(2\delta x)] = [\mathrm{T}(\delta x)]^2. \tag{18}$$

In practice, it is necessary to use high-precision during computation (e.g. double-precision in FORTRAN), and to ensure that energy conservation is preserved after each doubling. The new matrices may then be used to calculate the transfer matrices for distance $4\delta x$, $8\delta x$, and so on. A distance $2^n\delta x$ only requires the application of this 'matrix doubling' $n$ times. The spectrum of electrons and photons after a large distance $\delta x$ is then given by

$$[\mathrm{F}(x+\Delta x)] = [\mathrm{T}(\Delta x)][\mathrm{F}(x)] \tag{19}$$

where $[\mathrm{F}(x)]$ represents the input spectra, and $\Delta x = 2^n\delta x$. In this way, cascades over long distances can be modelled quickly and efficiently. For example the present calculation requires fewer than 40 matrix doublings.

### 3.4 Redshifting

The approach we adopt is to propagate over successive distances which would give rise to a change in $\log(1+z)$ equal to the width of the energy bins,

$$\Delta \log(1+z) = 0.1. \tag{20}$$

That is, if we are propagating from $z = 9$ ($\log[1+z] = 1.0$) to the present, we first propagate from $\log(1+z) = 1.0$ to $\log(1+z) = 0.9$ using matrices calculated at $\log(1+z) = 0.95$. In calculating the matrices, we assume that all components of the radiation field vary with redshift in the same way as the cosmic microwave background, i.e.

$$n(\varepsilon, z) = n\left(\frac{\varepsilon}{(1+z)}, z = 0\right)(1+z)^2. \tag{21}$$

This is not exact because it does not take account of the evolution of the infrared and radio backgrounds due to density or luminosity evolution of galaxies and AGN, and will introduce some error at large redshifts ($\sim 4 - 10$). The intergalactic magnetic field we use is obtained by assuming magnetic flux conservation per co-moving area, i.e. $B(z) = B_0(1+z)^2$ where $B_0$ is the magnetic field adopted for the present epoch. In this paper we will generally use $B_0 = 10^{-9}$ gauss which is the at the upper end of estimates of the average extragalactic field [39], but will also discuss lower fields. After propagation from $\log(1+z) = 1.0$ to $\log(1+z) = 0.9$ we redshift the energy bin contents of the vectors representing the particle spectra,

$$F_i^\alpha \leftarrow F_{i+1}^\alpha, \tag{22}$$

and then propagate from $\log(1+z) = 0.9$ to $\log(1+z) = 0.8$ using matrices calculated at $\log(1+z) = 0.85$, etc., redshifting the spectra and recalculating the matrices after each of these "big steps".

# 4 The origin of cosmic rays below the GZK cut-off

Cosmic rays with energies up to $10^{14}$ eV are thought to arise predominantly through shock acceleration by supernova remnants in our Galaxy [42], and this hypothesis has recently gained support from direct evidence for the first time [43, 44]. Slightly higher energies may be possible [45], but not high enough to explain the smooth extension of the spectrum apparently up to $10^{18}$ eV. Several explanations for the origin of the cosmic rays to this energy have been suggested: reacceleration of the supernova component while still inside the remnant [46] or by several supernovae exploding into a region evacuated by a pre-supernova star [47]; acceleration in shocks inside the strong winds from hot stars or groups of hot stars [48]; and a contribution from neutrons which escape from active galactic nuclei [49]. At $5 \times 10^{18}$ eV the spectral slope changes, and there is evidence for a lightening in composition [50] and it is likely this marks a change from galactic cosmic rays being dominant to extragalactic dominance.

The subject of the origin of cosmic rays above this energy up to $10^{20}$ eV has been reviewed extensively [51, 52, 53], and one of the very few plausible acceleration sites may be associated with the radio lobes of powerful radio galaxies, either in the hot spots [32] or possibly the cocoon or jet [54]. One-shot processes comprise another possible class of sources [55, 56]. Acceleration at the termination shock of the galactic wind from our Galaxy has been also been suggested [57], but is difficult to accept due to the lack of an observed anisotropy associated with the Galaxy. There are claims of a preferred phase in the Right Ascension distribution of cosmic rays above $10^{19}$ eV [58], but the magnitude of the anisotropy is almost consistent with fluctuations arising by chance [6]. However, a very recent re-evaluation of the world data set of cosmic rays has shown that there is a correlation of the arrival directions of cosmic rays above $4 \times 10^{19}$ with the supergalactic plane [59], lending support to an extragalactic origin above this energy, and in particular to models where "local" sources ($< 100$ Mpc) would appear to cluster near the supergalactic plane (e.g. powerful radio galaxies as in the model of Rachen and Biermann [32]).

Rachen and Biermann [32] have demonstrated that cosmic ray acceleration in Fanaroff-Riley Class II (FR II) radio galaxies can fit the observed spectral shape and the normalization at $10^{19}$ – $10^{20}$ eV to within a factor of less than 10. The predicted spectrum below this energy also fits the proton spectrum inferred from Fly's Eye data [60]. We have repeated the calculation of Rachen and Biermann in order to calculate the associated $\gamma$-ray and neutrino intensities. The source spectrum is assumed to be $E^{-2}$, and to extend up to a cut-off energy $E_c$; the cut off can either be sharp or exponential, $\exp(-E/E_c)$. We inject this spectrum at various redshifts up to $z = 9$ and propagate to $z = 0$ as described above, assuming $B_0 = 10^{-9}$ gauss, to obtain the spectra of all species considered. Throughout this paper we adopt $q_0 = 0.5$ and $H_0 = 75$ km s$^{-1}$ Mpc$^{-1}$ except as specified below. Dividing the proton spectrum by the injection spectrum, we obtain the "modification factor" $M(E, z)$ for injection at redshift $z$ as defined by Rachen and Biermann, and we show $M(E, z)$ for $E_c = 3 \times 10^{11}$ GeV from our present work in Fig. 7(a) for a sharp cut-off and in Fig. 7(b) for an exponential cut-off.

Because of the different techniques and assumptions used by different authors to propagate UHE protons, differences can arise in the predicted spectra. For example, Yoshida and Teshima [16] point out that structures of the bump and cut-off obtained by an analytic method are steeper than those obtained by a numerical/Monte Carlo method. These differences are most evident in the modification factors. We have added to Fig. 7(b) modification factors for an exponential cut-off used by Rachen and Biermann [32] which, apart from the spike, are generally in reasonable agreement with the present work. Figure 8 of Rachen and Biermann [32] and Figure 4 of Yoshida and Teshima [16] give modification factors for very similar inputs to those in Fig. 7(a), and we make a comparison between the three calculations in Fig. 8. Fig. 8(a) shows results for propagation over $\sim 256$ Mpc, Fig. 8(b) for $z \sim 0.6$, and Fig. 8(c) for $z = 1$. The differences in the modification factors between different calculations originate in the differences in the energy-loss distances evident in Fig. 4. For example, the peak at 10 EeV in the modification factor of Yoshida and Teshima [16] for $z = 0.5$ is probably attributable to their significantly higher energy-loss distance for pair production at these energies.

For an injection spectrum of $Q(E,z)$ protons per unit co-moving volume per unit energy per unit time, the intensity at Earth at energy $E$ is given by

$$I(E) = \frac{1}{4\pi}\int_{z_{min}}^{z_{max}} \mathrm{d}z\, M(E,z)\frac{(1+z)^2}{4\pi d_L^2}\frac{\mathrm{d}V_c}{\mathrm{d}z}Q((1+z)E,z) \tag{23}$$

where $d_L$ and $V_c$ are luminosity distance and co-moving volume. We consider an injection spectrum of the form

$$Q(E,z) = \frac{\Psi(z)E^{-2}g(E)}{\ln\gamma_c}, \tag{24}$$

i.e. an $E^{-2}$ spectrum which is modified at high energies by a cut-off function $g(E)$. The spectrum at Earth is then

$$I(E) = \frac{E^{-2}}{4\pi\ln\gamma_c}\int_{z_{min}}^{z_{max}} \mathrm{d}z\, g((1+z)E)M(E,z)\frac{\Psi(z)}{4\pi d_L^2}\frac{\mathrm{d}V_c}{\mathrm{d}z}, \tag{25}$$

where $\Psi(z)$ is the power per co-moving volume injected in cosmic ray protons. Rachen and Biermann have estimated the power per co-moving volume injected in cosmic ray protons in their model based on an analysis of observational data on powerful radio galaxies, and used radio luminosity functions of Peacock [61] which were obtained assuming $H_0 = 50$ km s$^{-1}$ Mpc$^{-1}$. We use $\Psi(z)/f$ from Figure 2 of Rachen and Biermann for radio luminosity function RLF2, where $f = 3.9$ is a "fudge factor" used by Rachen and Biermann, and calculate the cosmic ray intensity from Equation 25 taking $z_{min} = 0.03$, $z_{max} = 9$. We use $H_0 = 50$ km s$^{-1}$ Mpc$^{-1}$ and $q_0 = 0.5$ to calculate $\mathrm{d}V_c/\mathrm{d}z$ and $d_L$ (to be consistent with radio luminosity function used), and $M(E,z)$ for an exponential cut-off $g(E) = \exp(-E/E_c)$ from our Monte Carlo–matrix calculation for $H_0 = 75$ km s$^{-1}$ Mpc$^{-1}$. Our results for $E_c = 300$ EeV and 3000 EeV are shown in Figs. 9 and 10 for $H_0 = 75$ km s$^{-1}$ Mpc$^{-1}$ and $q_0 = 0.5$ where we have used the same fudge factors as Rachen and Biermann ($f = 3.9$ for 300 EeV and 3.5 for 3000 EeV). In part (a) we plot $E^{2.75}$ times the proton intensity and compare our results with those of Rachen and Biermann. We note that the agreement is very good given the differences in techniques noted earlier. In addition to nucleons, we also follow the secondary particles in the cascade, such as photons, electrons and neutrinos, and intensities of these particles are also shown in Figs. 9(a) and 10(a). The energy per decade going into secondaries is shown in Figs. 9(b) and 10(b) where we plot $E^2$ times the intensity. As we would expect, for the higher cut-off energy the $\gamma$-ray and neutrino intensities are much higher.

# 5 The origin of the highest energy cosmic rays

The two highest energy cosmic ray events are anomalous because they lie above the expected GZK cut-off and hence probably come from very nearby sources [14]. It is therefore difficult to explain the super-GZK events as an extension of the UHE cosmic ray spectrum produced in powerful distant radio galaxies. This conclusion is reinforced by the apparent lack of association with any likely extragalactic object, but we note that one intriguing possibility is that the highest energy cosmic rays might be associated with strong gamma ray burst sources [62, 63]. So it seems that either UHE cosmic rays are not accelerated as described in the previous section, or the super-GZK events belong to a new population; this possibility has been mentioned in the light of the paucity of events in the $1 - 2 \times 10^{20}$ eV range and hint of a turn-over above $5 \times 10^{19}$ eV [52].

Of course, the super-GZK events may result from a non-acceleration process. It has been suggested that they could be produced by the decay of supermassive X particles [64, 65, 66, 67], themselves radiated during collapse or annihilation of topological defects, remnants of an early stage in the evolution of the Universe. The X particles have GUT-scale masses of order $10^{15} - 10^{16}$ GeV, and decay into leptons and quarks at lower energies. The quarks themselves fragment into a jet of hadrons which, it is supposed, could produce the highest energy cosmic rays, although there is some debate as to whether a large enough fraction of the energy of the defect could end up in high energy particles [68]. In any case, much of the radiation is likely to emerge in the electromagnetic channel and one of the possibilities we shall explore in the next section is the idea that the super-GZK events could be due to an electromagnetic cascade initiated by such processes.

In an attempt to examine the question of where the super-GZK events may have originated, we first consider the injection of monoenergetic protons of energy $E_\delta = 3 \times 10^{20} - 10^{23}$ eV at various fixed distances: 8, 16, 32 Mpc, etc., up to $z = 9$. An example of the flux of protons and neutrons after propagation to Earth is given in Fig. 11. For each proton injection energy, we normalize to the observed intensity at $3 \times 10^{20}$ eV, and choose the injection distance that requires the least energy input. We then look at the predicted integral intensity of observable particles (nucleons, $\gamma$-rays, electrons) above $3 \times 10^{20}$ eV and compare it with the integral intensity corresponding to the $3 \times 10^{20}$ eV event. Three examples are given in Fig. 12(a)–(c); the first two are just allowed, but the third is ruled out by observations because of the enormous intensity above $3 \times 10^{20}$ eV, particularly of $\gamma$-rays.

In this way we find that for this type of model, injection energies above $3 \times 10^{12}$ GeV are ruled out, and for those models below this energy, only distances less than 30 – 50 Mpc are allowed. This probably rules out cosmological defect models in which the observed $3 \times 10^{20}$ eV events are due to production of hadrons near the GUT energy by a relatively small number of defects in the Universe. The predicted $\gamma$-ray intensity at lower energies may further constrain these models when new observations become available.

# 6 Primaries other than protons?

We now discuss the possibility that the highest energy cosmic rays are not single nucleons. Obvious candidates are heavier nuclei (e.g. Fe), $\gamma$-rays, and neutrinos. In general it is even more difficult to propagate nuclei than protons, because of the additional photonuclear disintegration which occurs [4, 3, 14]. Weakly interacting particles such as neutrinos will have no difficulty in propagating over extragalactic distances, of course. This possibility has been considered, and generally discounted [69, 14], mainly because of the relative unlikelihood of a neutrino interacting in the atmosphere, and the necessarily great increase in the luminosity required of cosmic sources.

The possibility that the super-GZK events are $\gamma$-rays has been discussed recently [69] and, although not completely ruled out, the $3 \times 10^{20}$ eV event has an air shower development profile which seems inconsistent with a $\gamma$-ray primary. Most of the emission by topological defects is likely to be deposited in electromagnetic cascades in the background radiation of primary energy up to $\sim 10^{14}$ GeV. If the super-GZK events are due to emission by topological defects one would expect $\gamma$-rays to make up most of the UHE intensity at Earth. Here we investigate the possibility that the UHE cosmic rays at Earth are in fact $\gamma$-rays produced by electromagnetic cascades initiated by $\gamma$-rays resulting from emission by, or decay of, topological defects. Such high energy electromagnetic cascades have been considered by several authors [70, 71, 64]. We consider injection of monoenergetic photons of energy $E_\delta = 10^{14}$ GeV at various fixed distances, 1, 2, 4, 8, 16, 32 Mpc, etc., up to $z = 9$. At the highest energies the electromagnetic cascade is very sensitive to the magnetic field because of the increasing importance of synchrotron losses. The flux at Earth is shown assuming $B = 10^{-9}$ gauss in Fig. 13(a) where the spectrum has been multiplied by $E^{2.75}$ (we will investigate lower $B$ below). The "hump" in the spectrum at $10^9$ GeV is due to synchrotron radiation in the $B = 10^{-9}$ gauss field by the first generation of electron-positron pairs (produced by double pair production on the microwave background or pair production on the radio background) which have energy $\sim 2 - 5 \times 10^{13}$ GeV. The flux above $10^{10}$ GeV is due to inverse-Compton emission, and the "valley" at $10^6$ GeV is due to photon-photon pair production on the microwave background. In Figure 13(b) we have normalized each of the curves in Figure 13(a) to the cosmic ray intensity at $3 \times 10^{20}$ eV, and it is obvious that we can conclude that a $\gamma$-ray origin of the super-GZK events from topological defects is ruled out on the basis of the associated lower energy radiation produced in the cascade, at least for an intergalactic magnetic field of $B = 10^{-9}$ gauss.

For lower magnetic fields, the situation may well be different. For example, if we choose a magnetic field a factor of $10^3$ lower we can move the hump into the valley and thereby reduce the intensity at 100 TeV – EeV energies. Whether such a low magnetic field is realistic is, however, debatable. In Figure 14(a) we show results for $B = 10^{-12}$ gauss and in Figure 14(b) we show results for $B = 3 \times 10^{-11}$ gauss normalized to the intensity at $3 \times 10^{20}$ eV. For $B = 3 \times 10^{-11}$ gauss all redshifts are ruled out because of the excessive intensity above $3 \times 10^{20}$ eV. For $B = 10^{-12}$ gauss, redshifts such that $\log(1 + z) < 0.2$ ($z < 0.6$) are ruled out because of the excessive intensity above $3 \times 10^{20}$ eV, and redshifts such that $\log(1+z) > 0.3$ ($z > 1$) are ruled out because of the excessive intensity below 100 GeV. It appears that for this magnetic field, it might be possible to explain the super-GZK events by $\gamma$-rays from topological defects at $z \sim 0.6 - 1$ as we are unable to rule this out at present. If defects at this redshift range are allowed by present data, future measurements at $\sim 100$ TeV [72] together with progress in determining the intergalactic magnetic field are most likely to constrain the models.

# 7 Conclusions

We have made a new calculation of the propagation of UHE cosmic rays through the extragalactic radiation fields over cosmological distances, and included in our work the calculation of secondary fluxes of $\gamma$-rays and neutrinos which result from interactions of the cosmic rays in the radiation field and subsequent cascading. We have repeated the important calculation of Rachen and Biermann [32] of the cosmic ray intensity resulting from acceleration in FR II radio galaxies, and find the accompanying intensities of $\gamma$-rays and neutrinos are not in conflict with observation.

In an attempt to explain the super-GZK events we have also considered injection of monoenergetic protons at various redshifts. We find that injection of protons of energy greater than $3 \times 10^{12}$ GeV at any redshift is ruled out because an unacceptably high intensity of observable particles (protons, neutrons, $\gamma$-rays) above $3 \times 10^{11}$ GeV would be produced. For injection energies less than $3 \times 10^{12}$ GeV, distances less than 30 – 50 Mpc are acceptable. We conclude from this that it is unlikely the super-GZK events could be explained by hadrons resulting from emission by, or decay of, topological defects.

For reasonable magnetic fields ($> 3 \times 10^{-11}$ gauss) injection of $\gamma$-rays at $\sim 10^{14}$ GeV, such as would arise in topological defect models, are ruled out by excessive $\gamma$-ray intensities at TeV and EeV energies, or above 300 EeV. For much lower magnetic fields (e.g. $10^{-12}$ gauss), it may be possible for this model to accommodate the data for injection at redshifts $z \sim 0.8$. In this case the predicted intensity of $\gamma$-rays at 100 TeV should be measurable and new measurements may be able to eliminate this scenario.

# 8 Acknowledgments

RJP thanks Peter Biermann and Jörg Rachen for helpful discussions and for providing their data for comparison used in Figs. 4, 7 and 8. We thank David Bird, Phil Edwards, Mikhail Entel and Greg Thornton for commenting on the original manuscript. This work is supported by a grant from the Australian Research Council.

## A Transfer matrices

If $\delta x$ is much shorter than the shortest effective interaction distance in the cascade, then

$$
\begin{aligned}
\mathrm{T}_{ij}^{\gamma\gamma}(\delta x) &\approx \delta_{ij}[1-\delta t\Gamma_{\gamma}(E_i)], & (26)\\
\mathrm{T}_{ij}^{\mathrm{e}\gamma}(\delta x) &\approx \delta t[\Gamma_{\mathrm{IC}}(E_i)Y_{ij}^{\mathrm{IC}\ \gamma}+\Gamma_{\mathrm{syn}}(E_i)Y_{ij}^{\mathrm{syn}\ \gamma}], & (27)\\
\mathrm{T}_{ij}^{\mathrm{p}\gamma}(\delta x) &\approx \delta t\Gamma_{\mathrm{phot}}(E_i)Y_{ij}^{\mathrm{phot}\ \gamma}, & (28)\\
\mathrm{T}_{ij}^{\mathrm{n}\gamma}(\delta x) &\approx \delta t\Gamma_{\mathrm{phot}}(E_i)Y_{ij}^{\mathrm{phot}\ \gamma}, & (29)\\
\mathrm{T}_{ij}^{\gamma\mathrm{e}}(\delta x) &\approx \delta t[\Gamma_{\mathrm{PP}}(E_i)Y_{ij}^{\mathrm{PP\ e}}+\Gamma_{\mathrm{DPP}}(E_i)Y_{ij}^{\mathrm{DPP\ e}}], & (30)\\
\mathrm{T}_{ij}^{\mathrm{ee}}(\delta x) &\approx \delta_{ij}[1-\delta t\Gamma_{\mathrm{e}}(E_i)]+\delta t[\Gamma_{\mathrm{IC}}(E_i)Y_{ij}^{\mathrm{IC\ e}}+\Gamma_{\mathrm{TPP}}(E_i)Y_{ij}^{\mathrm{TPP\ e}}+\Gamma_{\mathrm{syn}}(E_i)Y_{ij}^{\mathrm{syn\ e}}], & (31)\\
\mathrm{T}_{ij}^{\mathrm{pe}}(\delta x) &\approx \delta t[\Gamma_{\mathrm{phot}}(E_i)Y_{ij}^{\mathrm{phot\ e}}+\Gamma_{\mathrm{pair}}(E_i)Y_{ij}^{\mathrm{pair\ e}}], & (32)\\
\mathrm{T}_{ij}^{\mathrm{ne}}(\delta x) &\approx \delta t[\Gamma_{\mathrm{phot}}(E_i)Y_{ij}^{\mathrm{phot\ e}}+\frac{\delta t m_n c^2}{E_i\tau_n}Y_{ij}^{\mathrm{n\ dec\ e}}], & (33)\\
\mathrm{T}_{ij}^{\mathrm{pp}}(\delta x) &\approx \delta_{ij}[1-\delta t\Gamma_{\mathrm{p}}(E_i)]+\delta t[\Gamma_{\mathrm{phot}}(E_i)Y_{ij}^{\mathrm{phot\ p}}+\Gamma_{\mathrm{pair}}(E_i)Y_{ij}^{\mathrm{pair\ p}}], & (34)\\
\mathrm{T}_{ij}^{\mathrm{np}}(\delta x) &\approx \delta t\Gamma_{\mathrm{phot}}(E_i)Y_{ij}^{\mathrm{phot\ p}}+\delta_{ij}\frac{m_p}{m_n}\frac{\delta t m_n c^2}{E_i\tau_n}, & (35)\\
\mathrm{T}_{ij}^{\mathrm{pn}}(\delta x) &\approx \delta t\Gamma_{\mathrm{phot}}(E_i)Y_{ij}^{\mathrm{phot\ n}}, & (36)\\
\mathrm{T}_{ij}^{\mathrm{nn}}(\delta x) &\approx \delta_{ij}[1-\delta t\Gamma_{\mathrm{n}}(E_i)-\frac{\delta t m_n c^2}{E_i\tau_n}]+\delta t\Gamma_{\mathrm{phot}}(E_i)Y_{ij}^{\mathrm{phot\ n}}, & (37)\\
\mathrm{T}_{ij}^{\mathrm{p}\nu_{\mathrm{e}}}(\delta x) &\approx \delta t\Gamma_{\mathrm{phot}}(E_i)Y_{ij}^{\mathrm{phot}\ \nu_{\mathrm{e}}}, & (38)\\
\mathrm{T}_{ij}^{\mathrm{n}\nu_{\mathrm{e}}}(\delta x) &\approx \delta t\Gamma_{\mathrm{phot}}(E_i)Y_{ij}^{\mathrm{phot}\ \nu_{\mathrm{e}}}, & (39)\\
\mathrm{T}_{ij}^{\mathrm{p}\bar\nu_{\mathrm{e}}}(\delta x) &\approx [\delta t\Gamma_{\mathrm{phot}}(E_i)Y_{ij}^{\mathrm{phot}\ \bar\nu_{\mathrm{e}}}+\frac{\delta t m_n c^2}{E_i\tau_n}Y_{ij}^{\mathrm{n\ dec}\ \bar\nu_{\mathrm{e}}}], & (40)\\
\mathrm{T}_{ij}^{\mathrm{n}\bar\nu_{\mathrm{e}}}(\delta x) &\approx \delta t\Gamma_{\mathrm{phot}}(E_i)Y_{ij}^{\mathrm{phot}\ \bar\nu_{\mathrm{e}}}, & (41)\\
\mathrm{T}_{ij}^{\mathrm{p}\nu_{\mu}}(\delta x) &\approx \delta t\Gamma_{\mathrm{phot}}(E_i)Y_{ij}^{\mathrm{phot}\ \nu_{\mu}}, & (42)\\
\mathrm{T}_{ij}^{\mathrm{n}\nu_{\mu}}(\delta x) &\approx \delta t\Gamma_{\mathrm{phot}}(E_i)Y_{ij}^{\mathrm{phot}\ \nu_{\mu}}, & (43)\\
\mathrm{T}_{ij}^{\mathrm{p}\bar\nu_{\mu}}(\delta x) &\approx \delta t\Gamma_{\mathrm{phot}}(E_i)Y_{ij}^{\mathrm{phot}\ \bar\nu_{\mu}}, & (44)\\
\mathrm{T}_{ij}^{\mathrm{n}\bar\nu_{\mu}}(\delta x) &\approx \delta t\Gamma_{\mathrm{phot}}(E_i)Y_{ij}^{\mathrm{phot}\ \bar\nu_{\mu}}, & (45)\\
\mathrm{T}_{ij}^{\nu_e\nu_e}(\delta x) &= \mathrm{T}_{ij}^{\bar\nu_e\bar\nu_e}(\delta x)=\mathrm{T}_{ij}^{\nu_\mu\nu_\mu}(\delta x)=\mathrm{T}_{ij}^{\bar\nu_\mu\bar\nu_\mu}(\delta x)=\delta_{ij}, & (46)
\end{aligned}
$$

all other transfer matrices are null matrices, $c\delta t = \delta x$, and $\Gamma = c/\lambda_{\mathrm{eff}}$. The total interaction rates are

$$
\begin{aligned}
\Gamma_{\gamma}(E_i) &= \Gamma_{\mathrm{PP}}(E_i)+\Gamma_{\mathrm{DPP}}(E_i), & (47)\\
\Gamma_{\mathrm{e}}(E_i) &= \Gamma_{\mathrm{IC}}(E_i)+\Gamma_{\mathrm{TPP}}(E_i)+\Gamma_{\mathrm{syn}}(E_i), & (48)\\
\Gamma_{\mathrm{p}}(E_i) &= \Gamma_{\mathrm{phot}}(E_i)+\Gamma_{\mathrm{pair}}(E_i), & (49)\\
\Gamma_{\mathrm{n}}(E_i) &= \Gamma_{\mathrm{phot}}(E_i) & (50)
\end{aligned}
$$

and give the total interaction rates of photons, electrons, protons and neutrons. In the equations above, $\tau_n$ is the neutron's mean decay time, and we have used the following abbreviations: PP (photon-photon pair production), DPP (photon-photon double pair production), IC (inverse Compton), TPP (triplet pair production), pair (proton-photon pair production), phot (pion photoproduction), n dec (neutron decay), syn (synchrotron).

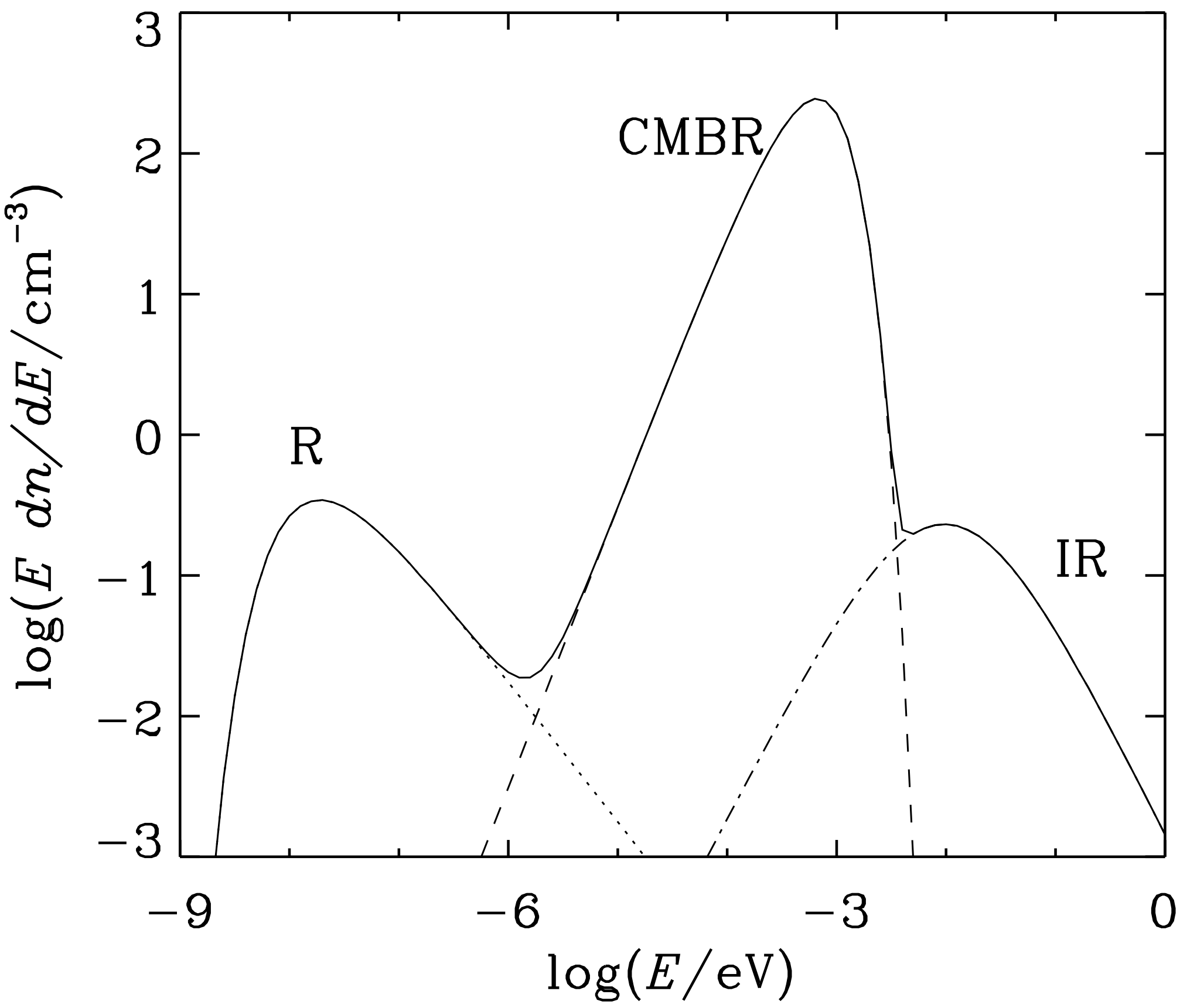

Figure 1: The composite extragalactic background photon spectrum, showing the contribution from the microwave blackbody (CMBR), the radio (R) and the infrared and optical (IR) extragalactic photon backgrounds.

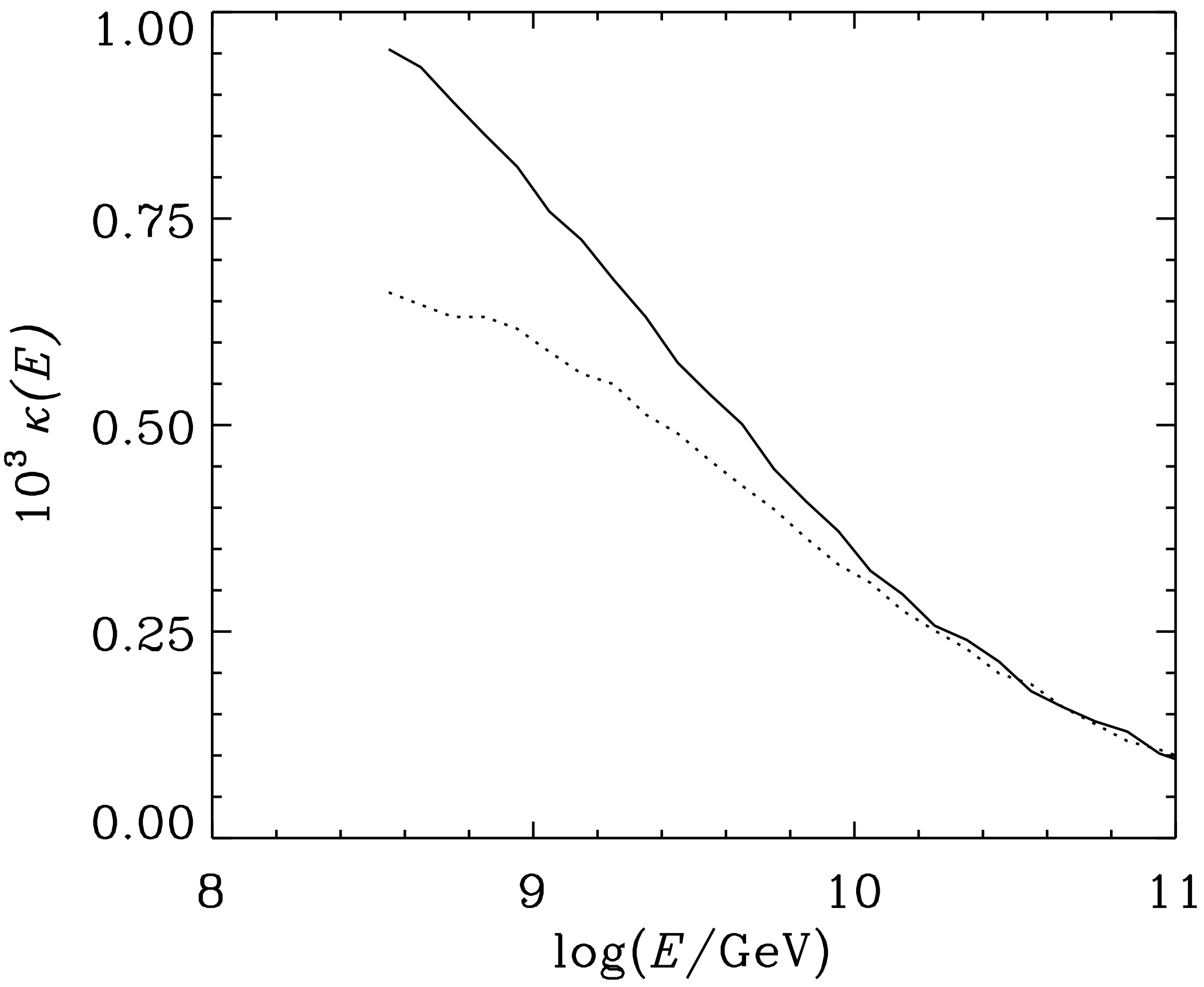

Figure 2: Mean inelasticity, $\kappa = \langle \Delta E \rangle / E$ for pair production calculated as described in the text using only the average positron lab frame energy (solid line) and using momnetum conservation in the proton rest frame, assuming no recoil, to obtain the electron energy (dotted lines).

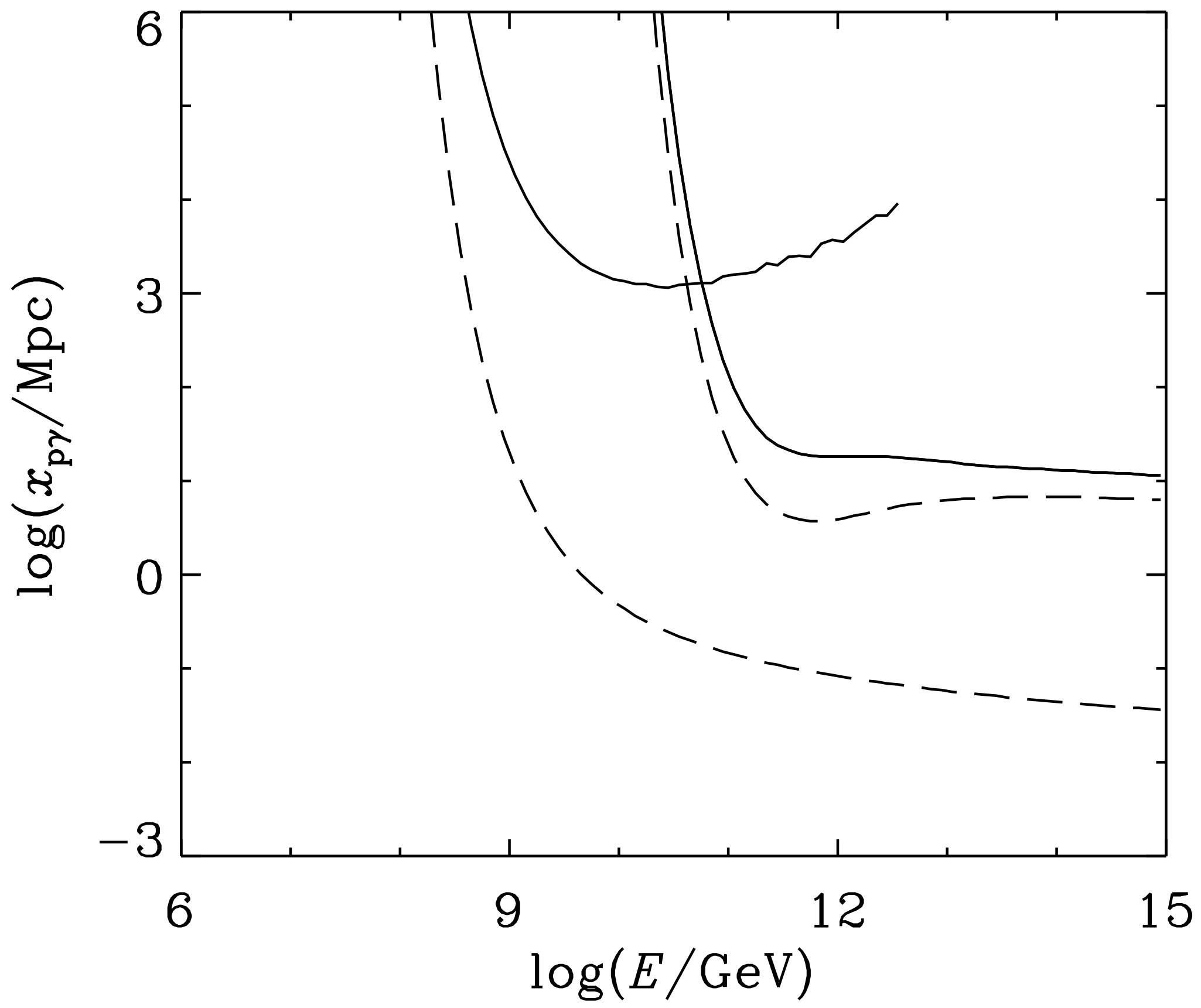

Figure 3: Mean interaction length (dashed lines) and energy-loss distance (solid lines), $E/(dE/dx)$, for proton-photon pair-production and pion-production in the microwave background (lower and higher energy curves respectively).

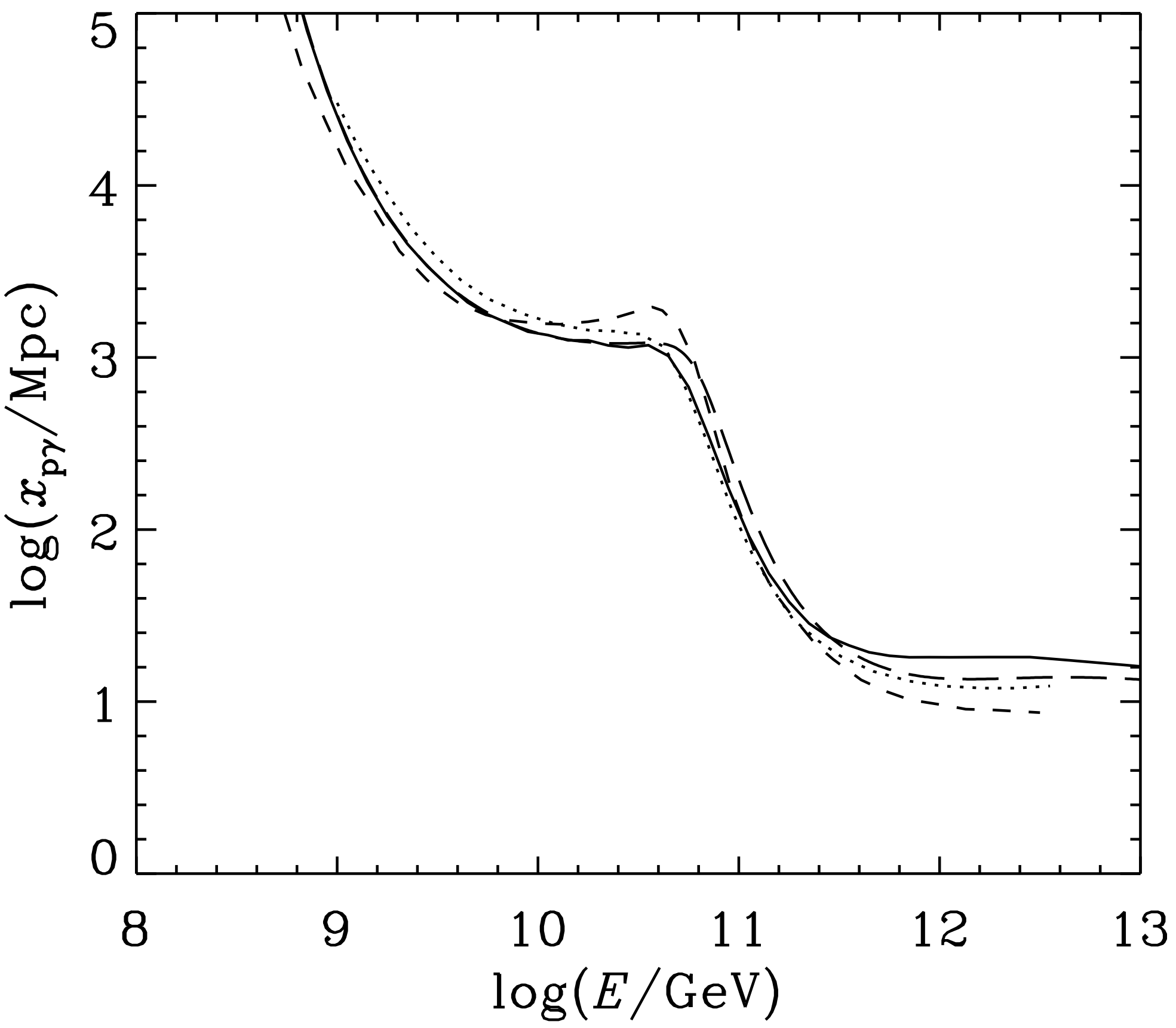

Figure 4: Total energy-loss distances in the microwave background radiation in the present work (solid line), and as calculated by Rachen and Biermann [32] (long-dashed line), Berezinsky and Grigor'eva [7] (dotted line), and Yoshida and Teshima [16] (short-dashed line).

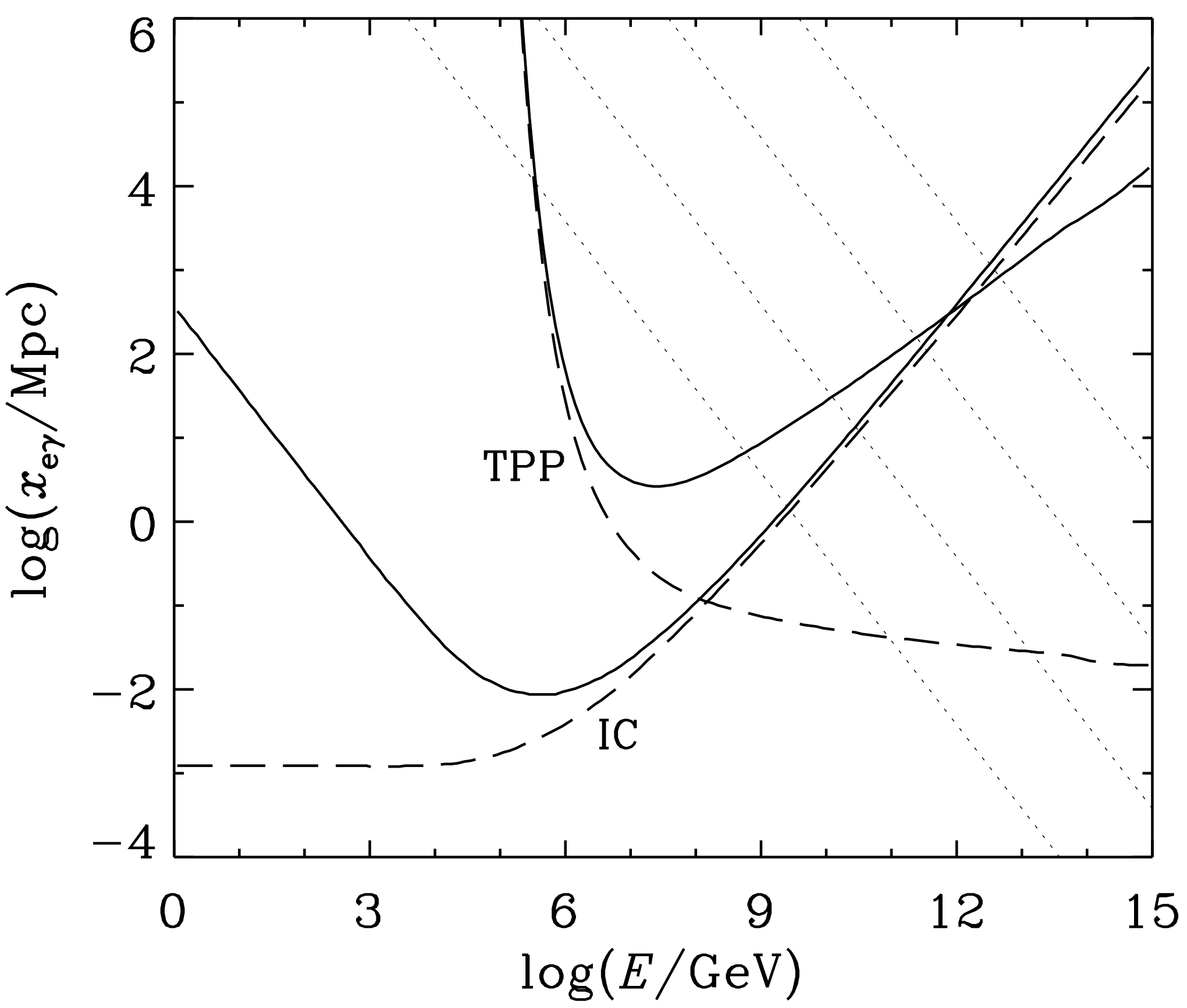

Figure 5: The mean interaction length (dashed line) and energy-loss distance (solid line), $E/(dE/dx)$, for electron-photon triplet pair production (TPP) and inverse-Compton scattering (IC) in the microwave background. The energy-loss distance for synchrotron radiation is also shown (dotted lines) for intergalactic magnetic fields of $10^{-9}$ (bottom), $10^{-10}$, $10^{-11}$, and $10^{-12}$ gauss (top).

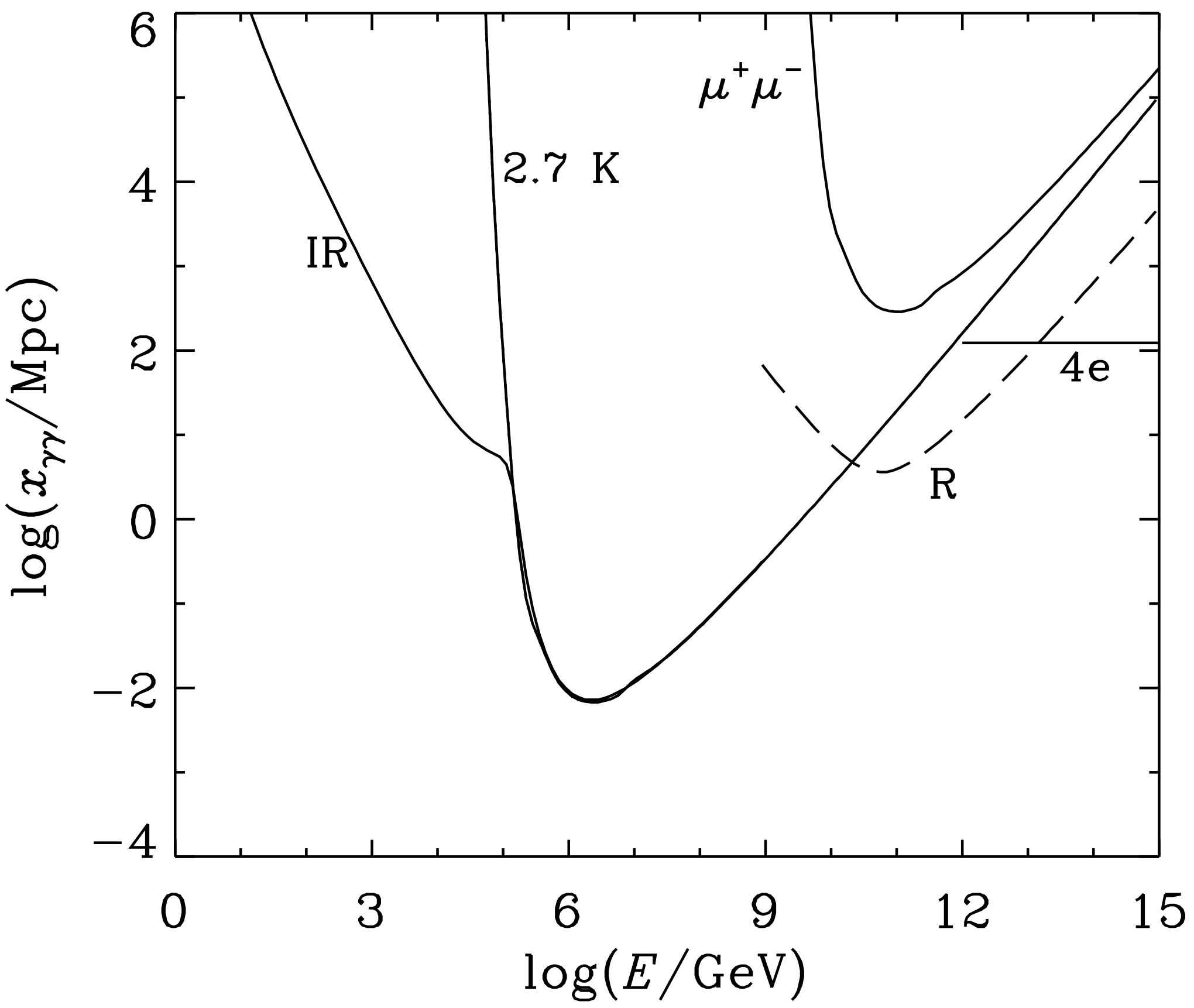

Figure 6: The mean interaction length for pair production for $\gamma$-rays in the microwave background (2.7K), the infrared and optical background (IR), and the Radio Background (R). Also shown are the mean interaction length for muon pair production ($\mu^+\mu^-$) and double pair production (4e) in the microwave background.

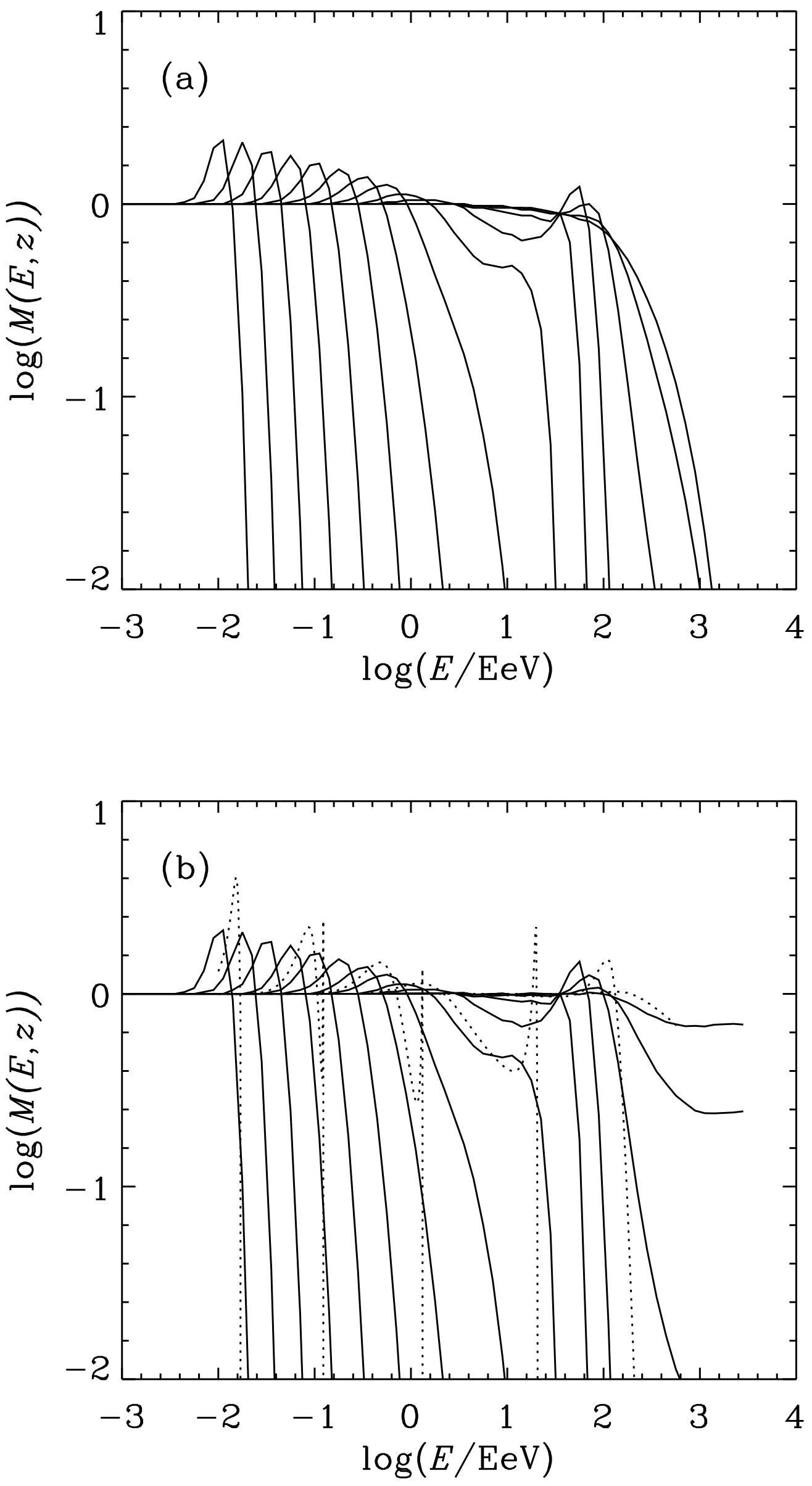

Figure 7: Modification factor, $M(E, z)$, for protons injected with (a) an $E^{-2}$ spectrum extending up to a sharp cut-off at $3 \times 10^{11}$ GeV and (b) a spectrum of the form $E^{-2} \exp(-E/E_c)$ with $E_c = 3 \times 10^{11}$ GeV. Solid curves are from the present work and are for (from right to left) distances 2, 8, 32, 128, 512 Mpc and $\log(1 + z) = 0.1, 0.2, 0.3, \ldots,$ 0.9 assuming $H_0 = 75$ km s$^{-1}$ Mpc$^{-1}$. Modification factors from the work of Rachen and Biermann [32] are also shown (dotted lines) in part (b) for (from right to left) distances 2 Mpc and 32 Mpc, and $\log(1 + z) = 0.1, 0.3, 0.6$ and 0.9.

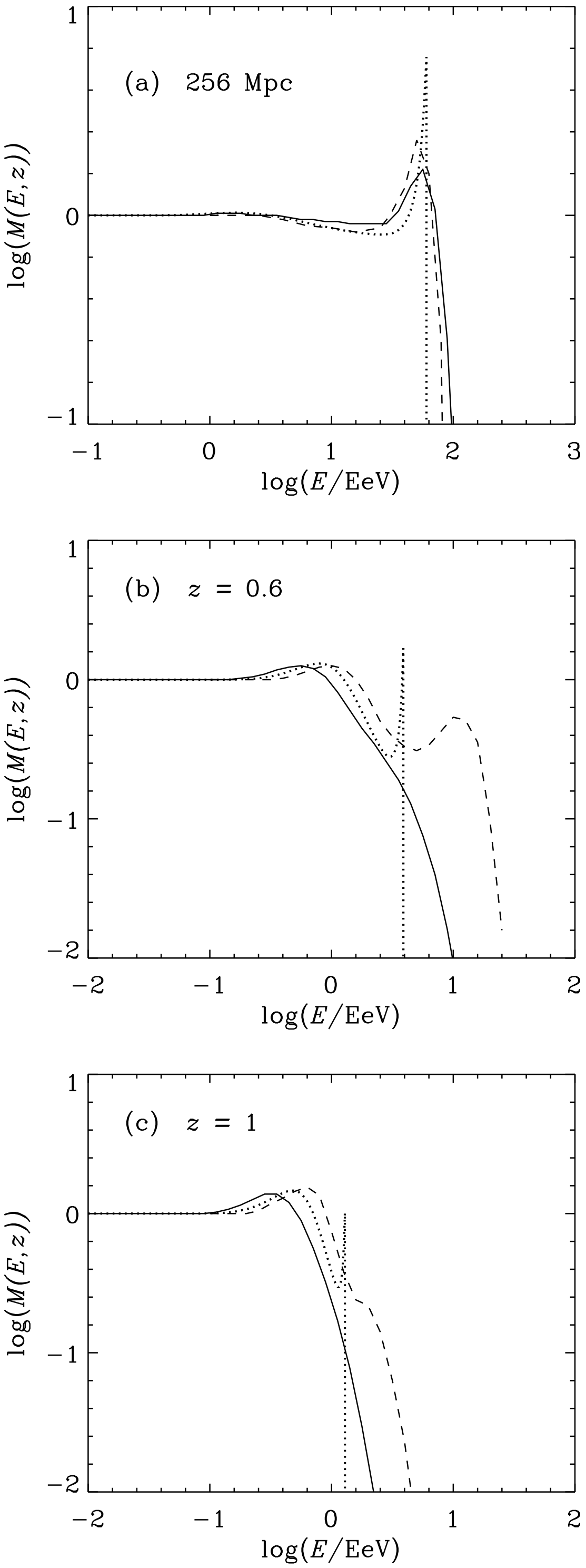

Figure 8: Comparison of modification factors for a sharp cut-off from the present work (solid lines) with those of Rachen and Biermann [32] (dotted lines) and Yoshida and Teshima [16] (dashed lines) for an $E^{-2}$ spectrum extending up to $3 \times 10^{11}$ GeV (present work, and Rachen and Biermann) or $10^{12}$ GeV (Yoshida and Teshima) for a source located at various distances: (a) 256 Mpc (present work), 240 Mpc (Rachen and Biermann), 228 Mpc (Yoshida and Teshima); (b) $z = 0.6$ (present work and Rachen and Biermann), $z = 0.5$ (Yoshida and Teshima); (c) $z = 1$ (all calculations). $H_0 = 75$ km s$^{-1}$ Mpc$^{-1}$ in all calculations.

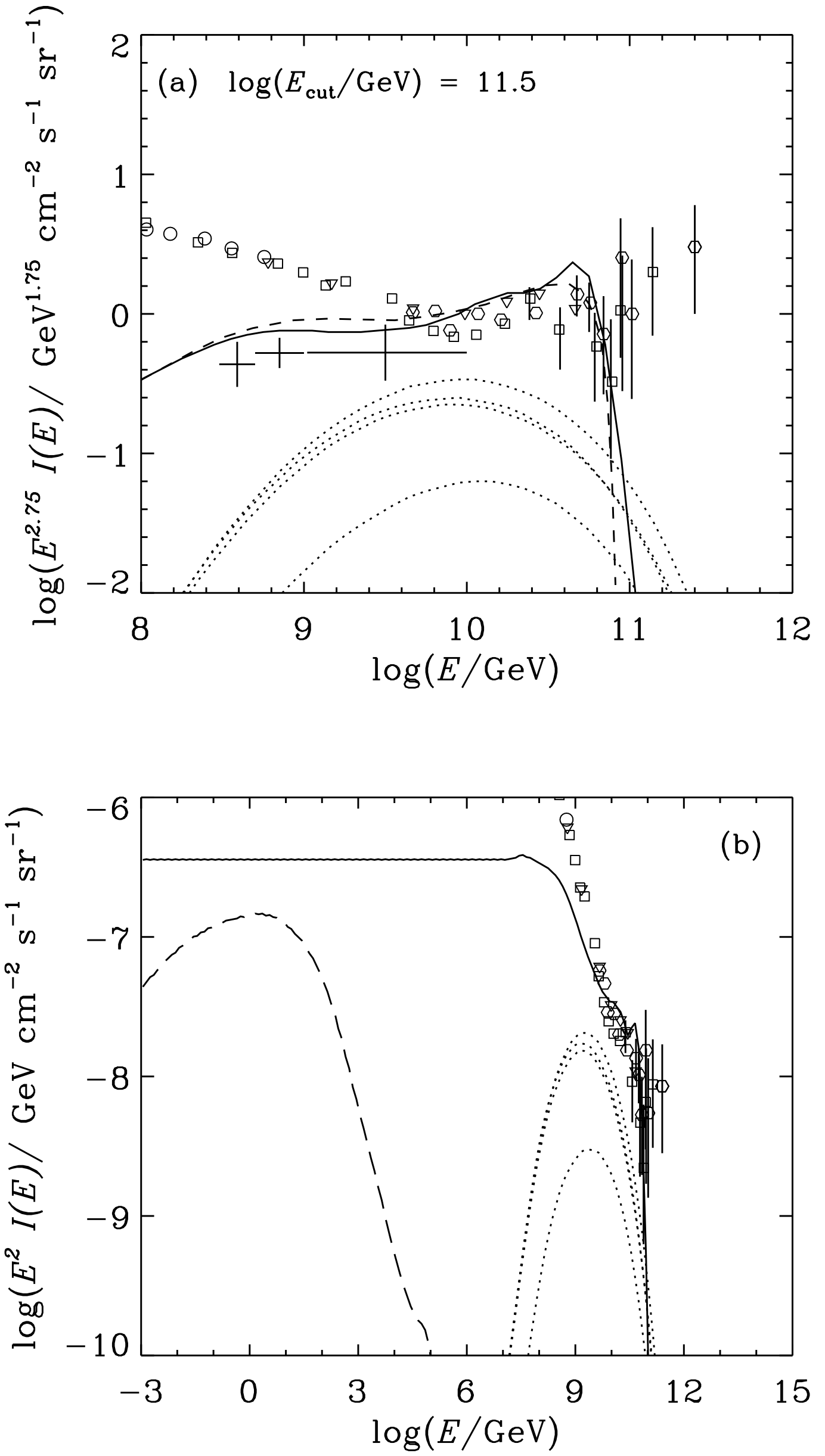

Figure 9: (a) Cosmic ray proton intensity multiplied by $E^{2.75}$ in the model of Rachen and Biermann for $H_0 = 75$ km s$^{-1}$ Mpc$^{-1}$ with proton injection up to $3 \times 10^{20}$ eV (solid line). Also shown are intensities of neutrinos (dotted lines, $\nu_\mu, \bar{\nu}_\mu, \nu_e, \bar{\nu}_e$ from top to bottom), and photons (long dashed lines). (b) As in part (a) but with intensities multiplied by $E^2$. The proton intensity calculated by Rachen and Biermann is shown in part (a) by the short dashed line. Data are from refs. [73] and [69]; large crosses at EeV energies are an estimate of the proton contribution to the total intensity based on Fly's Eye observations.

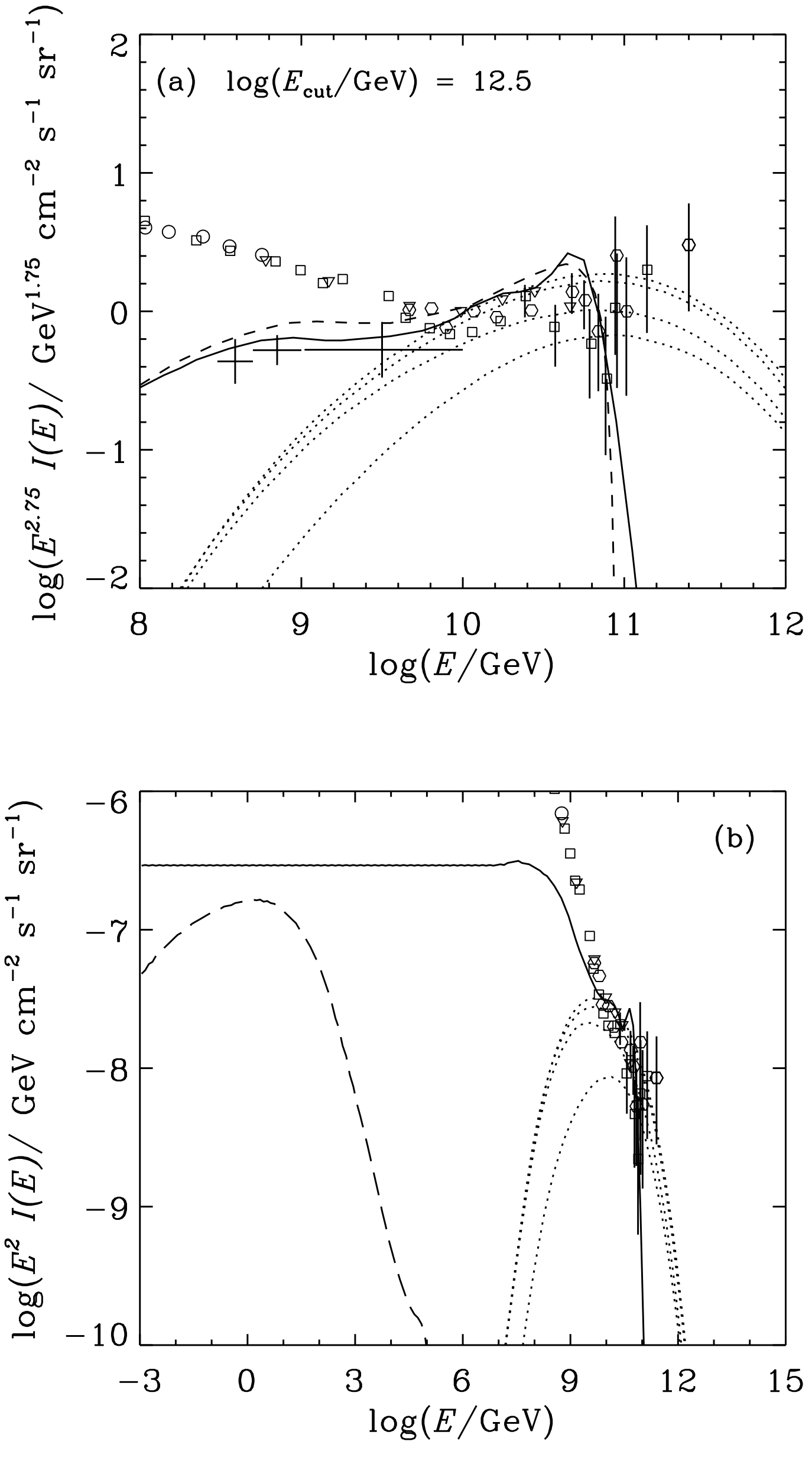

Figure 10: As Fig. 9, but for $E_c = 3 \times 10^{21}$ eV.

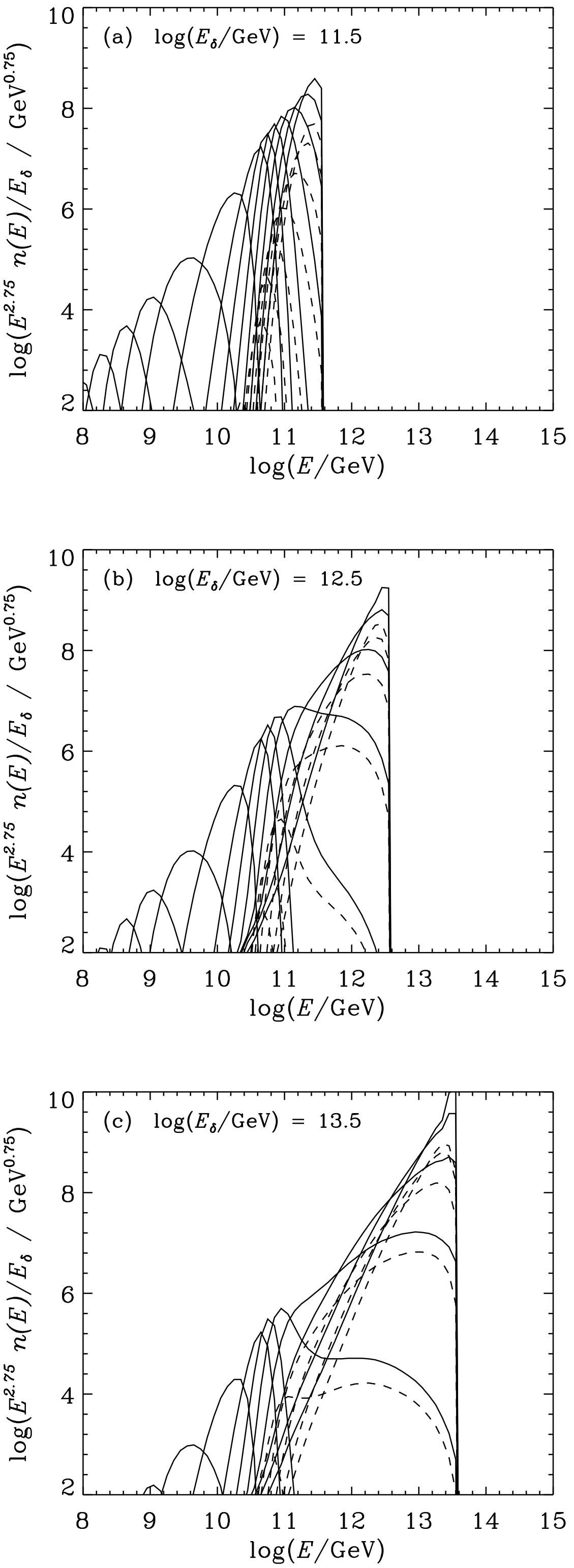

Figure 11: Cosmic ray nucleon spectrum due to monoenergetic injection of protons of energy $E_\delta =$ (a) $3 \times 10^{21}$ eV, (b) $3 \times 10^{22}$ eV, and (c) $3 \times 10^{23}$ eV, at distances of 2, 8, 32, 128, 512 Mpc, and $\log(1+z) = 0.1,\ 0.2$, etc. (solid lines: $p+n$, short dashed lines: $n$).

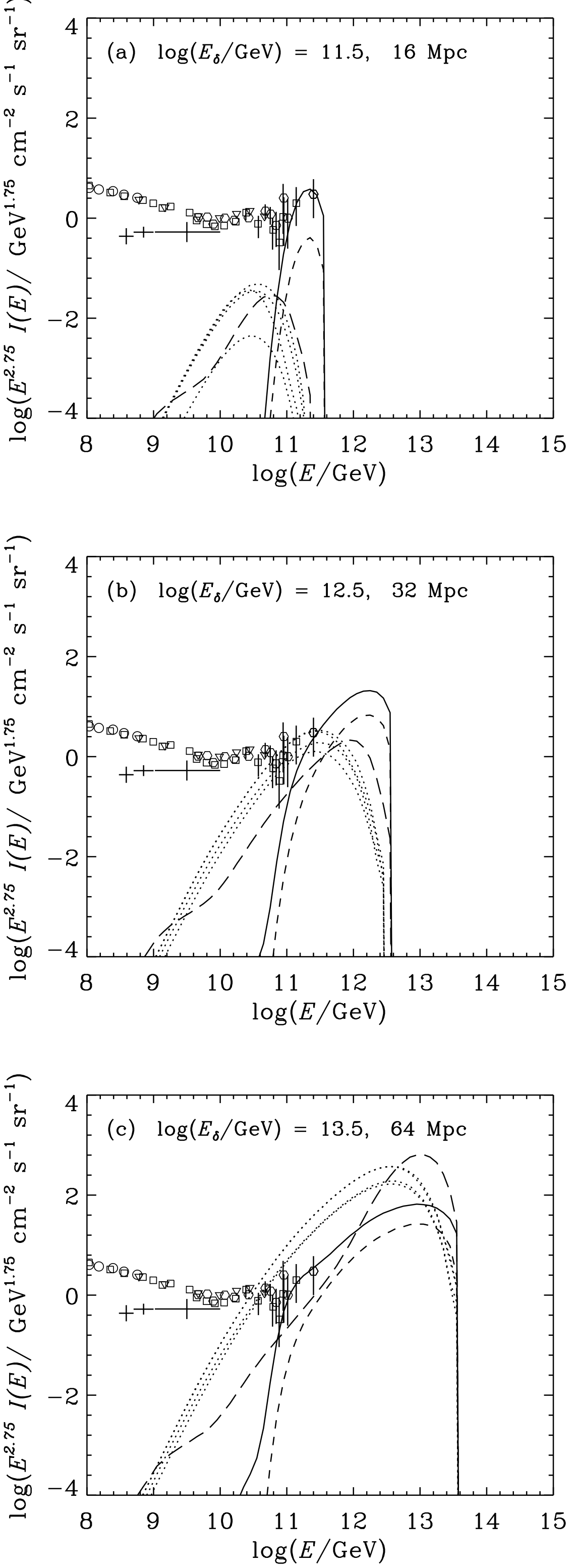

Figure 12: Spectrum from monoenergetic injection which best fits the intensity inferred from the observed $3 \times 10^{20}$ eV event for $E_\delta$ = (a) $3 \times 10^{21}$ eV, (b) $3 \times 10^{22}$ eV, and (c) $3 \times 10^{23}$ eV. Solid lines $p + n$; short-dashed lines $n$; dotted lines, $\nu_\mu, \bar{\nu}_\mu, \nu_e, \bar{\nu}_e$ from top to bottom; long-dashed lines photons.

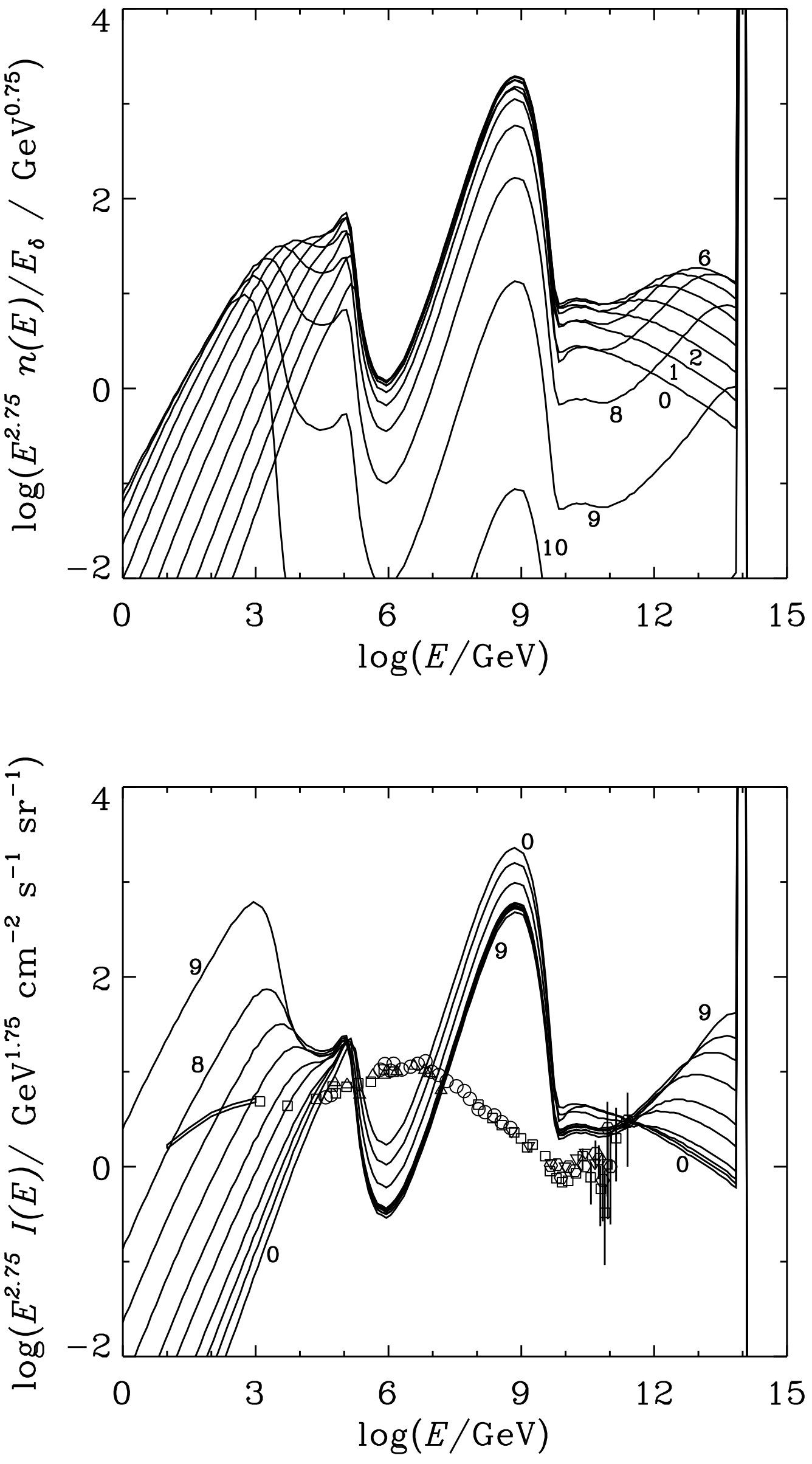

Figure 13: (a) The upper plot shows the spectrum of $\gamma$-rays which results after injection of a primary $\gamma$-ray of energy $E_\delta = 10^{14}$GeV for an intergalactic magnetic field of $10^{-9}$ gauss and subsequent cascading over distances $2^n$ Mpc, where $n$ is the number attached to the curve. (b) The lower plot shows the result of normalizing each curve in part (a) to the data at $3 \times 10^{20}$ eV.

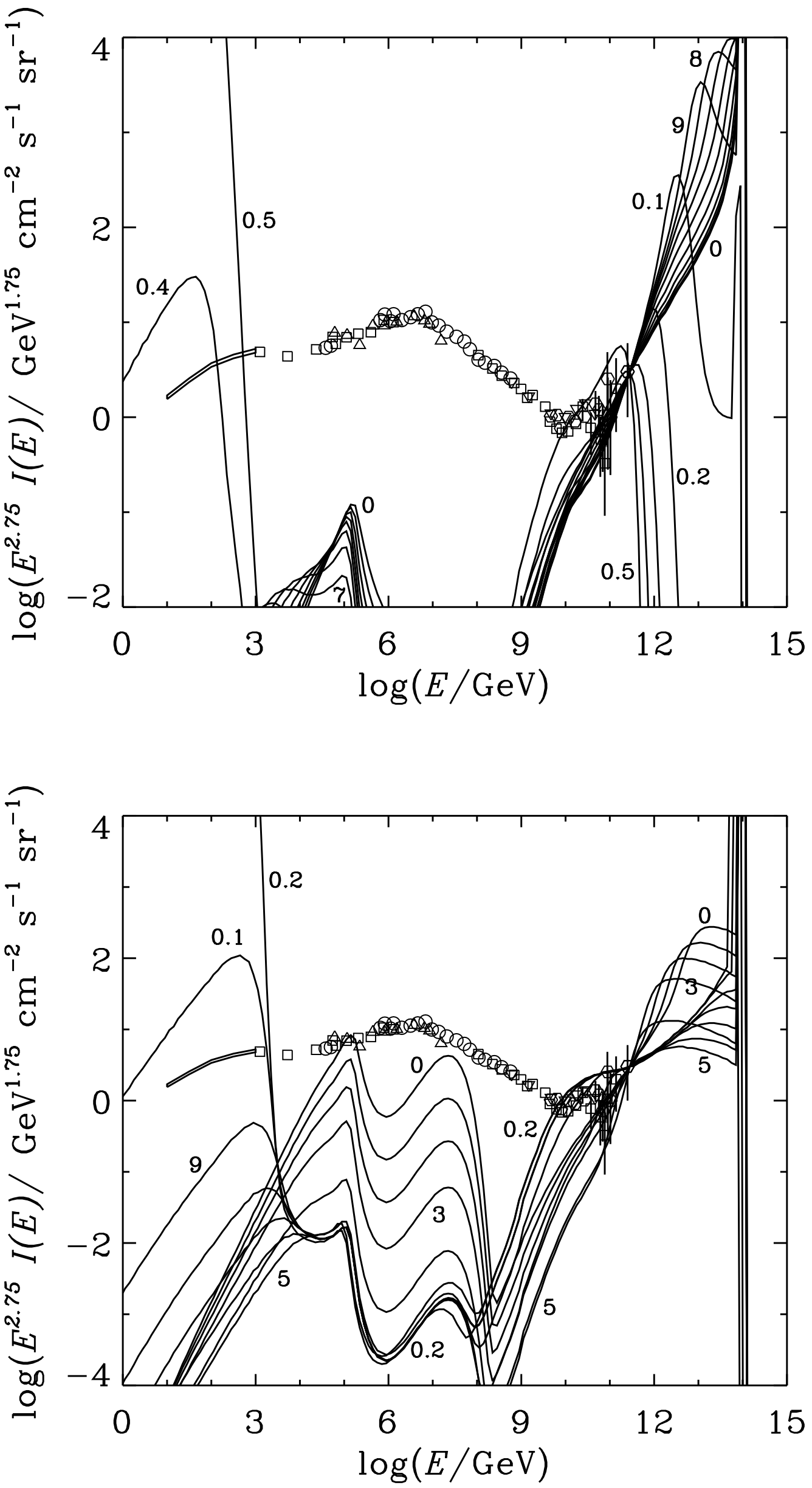

Figure 14: The intensity of $\gamma$-rays which results at Earth, after injection of energy $E_\delta = 10^{14}$ GeV for various distances (see Fig. 13) and normalized to the data at $3 \times 10^{20}$ eV for (a) upper plot – an intergalactic magnetic field of $10^{-12}$ gauss, and (b) lower plot – an intergalactic magnetic field of $3 \times 10^{-11}$ gauss.